\documentclass[10pt,twocolumn]{article}
\usepackage{fullpage,amsmath}

\newcommand{\ket}[1]{\mbox{$| #1 \rangle$}}  
\newcommand{\mb}[1]{\mbox{$#1$}}             

\newtheorem{thm}{Theorem}
\newtheorem{lem}{Lemma}
\newenvironment{proof}{\noindent {\bf Proof. }}{\hfill$\rule{.7ex}{.7ex}$}

\begin{document}
\title{A quantum algorithm for finding the minimum\thanks{{This work was supported by the {\sc alcom-it} Research 
        Programme of the EU, the {\sc rand2} Esprit Working Group, 
        and the ISI Foundation.}}}
\author{Christoph D\"urr\thanks{Laboratoire de Recherche en Informatique, 
        Universit\'e Paris-Sud, b\^at.\ 490, F--91405 Orsay, France.  Email: durr@lri.fr. } \and
Peter H{\o}yer\thanks{Dept.\ of Math.\ and Comp.\ Science, Odense 
        University, Campusvej~55, DK--5230 Odense M, Denmark. Email: u2pi@imada.ou.dk.  }}

\date{July 1996}
\maketitle

\section{Introduction}
Let~$T[0..N-1]$ be an unsorted table of $N$~items, each holding a
value from an ordered set.  For simplicity, assume that all values are
distinct.  The minimum searching problem is to find the index~$y$ such
that $T[y]$ is minimum. This clearly requires a linear number of
probes on a classical probabilistic Turing machine.

Here, we~give a simple quantum algorithm which solves the problem using
\mb{{\mathcal O}(\sqrt N)} probes. The main subroutine is the quantum
exponential searching algorithm of~\cite{BBHT96}, which itself is a
generalization of Grover's recent quantum searching
algorithm~\cite{Grover96}.  Due to a general lower bound
of~\cite{BBBV95}, this is within a constant factor of the optimum.

\section{The algorithm}
Our algorithm calls the quantum exponential searching algorithm
of~\cite{BBHT96} as a subroutine to find the index of a smaller item
than the value determined by a particular {\em threshold index}. The
result is then chosen as the new threshold. This process is repeated
until the probability that the threshold index selects the minimum is
sufficiently large.

If~there are $t\geq 1$ marked table entries, the quantum exponential
searching algorithm will return one of them with equal probability after
an expected number of ${\mathcal O}(\sqrt{N/t\,})$ {\em iterations}.
If~no entry is marked, then it will run forever. We obtain the following
theorem.

\begin{thm}						\label{thm}
The algorithm given below finds the index of the minimum value with
probability at least~$\frac12$. Its running time is ${\mathcal O}(\sqrt
N)$.
\end{thm}

We~first give the minimum searching algorithm, then the proof of the
probability of success.

\paragraph*{\sc Quantum Minimum Searching Algorithm}
\begin{enumerate}
\item 							\label{choose}
	Choose threshold index $0\leq y \leq N-1$ uniformly at random.
\item 							\label{rek}
	Repeat the following and interrupt it when the total running
	time is more than $22.5\sqrt N + 1.4\lg^2 N$.%
\footnote{As notation, we use $\lg$ for the binary logarithm and
$\ln$ for the natural logarithm.}
      	Then go to stage~\ref{rek}(\ref{obs}).
\begin{enumerate}
\item 							\label{init}
	Initialize the memory as $\sum_j \frac1{\sqrt{N}}
      	\ket{j} \ket{y}$.  \\
        Mark every item~$j$ for which $T[j]<T[y]$.
\item 							\label{search}
	Apply the quantum exponential searching algorithm
      	of~\cite{BBHT96}. 
\item 							\label{obs}
	Observe the first register: let~$y'$ be the outcome.  
      	If~$T[y']<T[y]$, then set threshold index~$y$ to~$y'$.
\end{enumerate}
\item 							\label{return}
	Return~$y$.
\end{enumerate}
\medskip

By convention, we assume that stage~\ref{rek}(\ref{init}) takes $\lg(N)$
time steps and that one iteration in the exponential searching algorithm
takes one time step. The work performed in the stages~\ref{choose},
\ref{rek}(\ref{obs}), and \ref{return} is not counted.

For the analysis of the probability of success, assume that there is no
time-out, that is, the algorithm runs long enough to find the
minimum. We refer to this as the {\em infinite algorithm}.  We~start by
analyzing the expected time to find the minimum.

At any moment the infinite algorithm searches the minimum among the
$t$~items which are less than~$T[y]$. During the execution any such
element will be chosen at some point as the threshold with some
probability. The following lemma states that this probability is the
inverse of the rank of the element and is independent of the size of the
table.

\begin{lem}                                   	\label{1r}
Let $p(t,r)$ be the probability that the index of the element of
rank~$r$ will ever be chosen when the infinite algorithm searches among
$t$~elements. Then $p(t,r) = 1/r$ if $r\leq t$, and $p(t,r)=0$
otherwise.
\end{lem}

\begin{proof}
The case $r>t$ is obvious. The case $r \leq t$ is proven by induction
on~$t$ for a fixed~$r$: The basis $p(r,r)=1/r$ is obvious since the
output distribution of the exponential searching algorithm is uniform.
Assume that for all $k\in[r,t]$ the equation $p(k,r)=1/r$ holds. Then
when $t+1$~elements are marked, $y$~is chosen uniformly among all
$t+1$~indices, and can hold an item of rank either~$r$, greater
than~$r$, or less than~$r$. Only the former two cases contribute to the
summation:
\begin{align*}
	 p(t+1,r) =& \frac1{t+1} + \sum_{k=r+1}^{t+1} \frac1{t+1} p(k-1,r)\\
   =& \frac1{t+1} + \frac{t+1-r}{t+1}\cdot \frac1r \\=& \frac1r\,. 
\end{align*}
\end{proof}

The expected number of iterations used by the exponential searching
algorithm of~\cite{BBHT96} to find the index of a marked item ---
among $N$~items where $t$~items are marked --- is at most
$\frac{9}{2}\sqrt{N/t\,}$ (see~\cite{BBHT96}).  We~can now deduce the expected
time used to find the minimum:

\begin{lem}                           		\label{exp}
The expected total time used by the infinite algorithm before~$y$ holds
the index of the minimum is at most 
\mb{m_0 = \frac{45}{4} \sqrt N + \frac{7}{10} \lg^2 N}.
\end{lem}

\begin{proof} The expected total number of time steps of 
stage~\ref{rek}(\ref{search}) before $T[y]$ holds the minimum is at most
\begin{align*}
\sum_{r=2}^N p(N,r) \frac{9}{2} \sqrt\frac N{r-1}
=&    \frac{9}{2} \sqrt N \sum_{r=1}^{N-1} \frac1{r+1} \frac1{\sqrt{r}} \\ 
\leq& \frac{9}{2} \sqrt N \left(\frac12 + \sum_{r=2}^{N-1} 
       {r^{-3/2}}\right)\\ 
\leq& \frac{9}{2} \sqrt N \left(\frac12 + \int_{r=1}^{N-1} 
       {r^{-3/2}} \mathrm d r\right)\\ 
=&    \frac{9}{2} \sqrt N \left(\frac12 + 
       \biggl[-2 {r^{-1/2}}\biggr]_{r=1}^{N-1}\right)\\
\leq& \frac{45}{4} \sqrt N.
\end{align*}

The expected total number of time steps of stage~\ref{rek}(\ref{init})
before $T[y]$ holds the minimum is at most
\begin{align*}
 \sum_{r=2}^N p(N,r) \lg N =& (H_N - 1) \lg N \\\leq& 
    \ln N \, \lg N \\ \leq& \frac{7}{10} \lg^2 N, 
\end{align*}
where $H_N$ denotes the~$N$th harmonic number. 
\end{proof}

Theorem~\ref{thm} follows immediately from lemma~\ref{exp} since after
at most \mb{2 m_0} iterations, $T[y]$ holds the minimum value with
probability at least~$\frac12$.

\section{Final remarks}
The probability of success can be improved by running the algorithm
$c$~times. Let~$y$ be such that $T[y]$ is the minimum among the
outcomes.  With probability at least $1-1/2^c$, $T[y]$ holds the minimum
of the table. Clearly, the probability of success is even better if we
run the algorithm only once with time-out \mb{c 2 m_0}, because then we
use the information provided by the previous steps.

If~the values in~$T$ are not distinct, we use the same algorithm as
for the simplified case.  The analysis of the general case is
unchanged, except that in lemma~\ref{1r}, the equation \mb{p(t,r) =
1/r} now becomes an inequality, \mb{p(t,r) \leq 1/r}.  Hence, the
lower bound for the success probability given in theorem~\ref{thm} 
for the simplified case is also a lower bound for the general case.

\section*{Acknoledgment}
We~wish to thank Stephane Boucheron for raising this problem to us and
also Richard Cleve and Miklos Santha for helpful discussions.

\end{document}